# First-principles investigation of Ag-, Co-, Cr-, Cu-, Fe-, Mn-, Ni-, Pd- and Rh-hexaaminobenzene 2D metal-organic frameworks

Bohayra Mortazavi[*,1], Masoud Shahrokhi[1], Meysam Makaremi[2], Gianaurelio Cuniberti[3,4,5] and Timon Rabczuk[6]

[1]*Institute of Structural Mechanics, Bauhaus-Universität Weimar, Marienstr. 15, D-99423 Weimar, Germany.*

[2]*Department of Materials Science and Engineering, University of Toronto, 184 College Street, Suite 140, Toronto, ON M5S 3E4, Canada.*

[3]*Institute for Materials Science, TU Dresden, 01062 Dresden, Germany*

[4]*Dresden Center for Computational Materials Science, TU Dresden, D-01062 Dresden, Germany*

[5]*Center for Advancing Electronics Dresden, TU Dresden, 01062 Dresden, Germany*

[6]*Department of Computer Engineering, College of Computer and Information Sciences, King Saud University, Riyadh, Saudi Arabia.*

**Abstract**

Hexaaminobenzene (HAB)-derived two-dimensional metal–organic frameworks (MOFs) *(Nature Energy 3(2018), 30-36)* have most recently gained remarkable attentions as a novel class of two-dimensional (2D) materials, with outstanding performances for advanced energy storage systems. In the latest experimental advances, Ni-, Co- and Cu-HAB MOFs were synthesized in 2D forms, with high electrical conductivities and capacitances as well. Motivated by these experimental advances, we employed first-principles simulations to explore the mechanical, thermal stability and electronic properties of single-layer Ag-, Co-, Cr-, Cu-, Fe-, Mn-, Ni-, Pd- and Rh-HAB MOFs. Theoretical results reveal that Co-, Cr-, Fe-, Mn-, Ni-, Pd- and Rh-HAB nanosheets exhibit linear elasticity with considerable tensile strengths. Ab-initio molecular dynamics results confirm the high thermal stability of all studied nanomembranes. Co- and Fe-HAB monolayers show metallic behavior with low spin-polarization at the Fermi level. Single-layer Ag-, Cu-, Cr-, and Mn-HAB however yield perfect half-metallic behaviors, thus can be promising candidates for the spintronics. In contrast, Ni-, Pd- and Rh-HAB monolayers exhibit nonmagnetic metallic behavior. The insights provided by this investigation confirm the stability and highlight the outstanding physics of transition metal-HAB nanosheets, which are not only highly attractive for the energy storage systems, but may also serve for other advanced applications, like spintronics.

Corresponding authors: *bohayra.mortazavi@gmail.com, #timon.rabczuk@uni-weimar.de;

# 1. Introduction

Design and fabrication of advanced energy storage systems, with improved reliability and power density, and faster charging/discharging rates, are currently among the key factors for the success in highly competitive technologies, like; mobile electronics and electric vehicles. Such a continues demand drives the progress of new materials that can improve the efficiency of energy storage systems. Two-dimensional (2D) materials have recently gained remarkable attentions for a wide-range of applications, from post-silicon nanoelectronics, to advanced energy storage systems and structural components. The interests toward the 2D materials was raised, because of the great success of graphene [1,2], with a unique combination of ultrahigh mechanical [3] and thermal conduction [4,5] properties outperforming all-known materials and very attractive optical and electronic characteristics [6–8]. In particular, 2D materials owing to their large surface area, flexibility, desirable mechanical and thermal stability, good electron mobility and fast ionic conductivity, have recently garnered remarkable attention for the application in advanced energy storage systems, such as the rechargeable metal-ion batteries and electrochemical supercapacitors [9–18].

Among the various families of 2D materials, nanomembranes with porous atomic lattices, like graphdiyne [19] structures with hybrid sp and $sp^2$ covalently bonded carbon atoms arranged in various crystalline, exhibit outstanding performances for energy storage applications [20–23]. Worthy to note that, Matsuoka *et al.* [24] in 2017 reported the first synthesize of crystalline graphdiyne nanosheets at a gas/liquid or liquid/liquid interface. In contrast with densely packed 2D materials like graphene and $MoS_2$, nanosheets with porous atomic lattices like graphdiyne provide easy access to built-in active sites for the electrochemical reactions and metal-ions adsorptions, which are the key factors to enhance the efficiency. On the other side, organic materials are considered among the most promising candidates for energy storage devices, owing to their redox activity, ubiquity and light weight [25]. Metal-organic frameworks (MOFs) in 2D forms with porous atomic lattices, show one of the most effective platforms for the next generation energy storage systems. In an exciting most recent experimental study by Feng *et al.* [26], hexaaminobenzene (HAB)-derived, porous and 2D MOFs were designed and fabricated experimentally. Cu- and Ni-HAB nanosheets were found to exhibit outstanding prospects for the application as supercapacitors, stemming from their excellent chemical stability in both acidic and basic aqueous solutions, high volumetric capacitances and high areal capacitances, highly reversible redox behaviours and good cycling stability [26]. In another latest experimental advance, Co-HAB MOF nanomembranes were for the first time experimentally realized by Park *et al.* [27]. Interestingly, Co-HAB nanosheets were confirmed to present promising performances a new electrode material for sodium-ion storage, with a high electrical conductivity and extremely high rate capability that outperforms many other organic and inorganic electrode materials for high-power Na-ion storage devices [27]. These latest experimental successes [26,27] in the fabrication of Co-, Cu- and Ni-HAB nanomembranes, raise the importance of theoretical studies in order to provide understanding of the intrinsic material properties. Now-a-days, theoretical techniques provide the unique opportunities for the researchers to examine new compositions and estimate properties and suggest possible synthesis routes [28–33]. Motivated by these experimental advances, in this investigation we employed first-principles density functional theory simulations to explore the mechanical, thermal stability and electronic properties of single-layer Ag-, Co-, Cr-, Cu-, Fe-, Mn-, Ni-, Pd- and Rh-HAB MOFs. The acquired results confirm the stability and reveal very attractive properties of this novel class of 2D materials and will hopefully motivate further theoretical and experimental studies.

## 2. Computational methods

Spin polarized density functional theory calculations in this work were conducted using the *Vienna Ab-initio Simulation Package* (VASP) [34–36] within the Perdew-Burke-Ernzerhof (PBE) functional [37] for the exchange correlation potential. The core electrons are replaced by Projected augmented wave (PAW) method [38] and pseudo-potential approach. A plane-wave cutoff energy of 500 eV was set for the DFT calculations. VESTA [39] package was employed for the graphical presentation of atomic structures. In order to acquire the energy minimized structures, the size of the unit-cell was uniformly changed and then the conjugate gradient method was employed for the geometry optimizations using a 3×3×1 Monkhorst-Pack [40] k-point mesh size. In this case, criteria for the convergence of the energy of electronic self consistence-loop was considered to be $10^{-4}$ eV and for the structural relaxation, the Hellmann–Feynman forces on each atom was taken to be 0.01 eV/Å. To evaluate the electronic band-structure, we used a denser k-point mesh size of 11×11×1. Mechanical properties were evaluated by performing uniaxial tensile simulations [32]. In order to examine the thermal stability of considered structures, ab-initio molecular dynamics (AIMD) simulations were performed for the hexagonal unit-cells, using the Langevin thermostat with a time step of 1 fs and 2×2×1 k-point mesh size [33].

## 3. Results and discussions

In Fig.1, energy minimized Ag-, Co-, Cr-, Cu-, Fe-, Mn-, Ni-, Pd- and Rh-HAB monolayers with hexagonal atomic lattices are illustrated. These nanosheets are consisting of a redox-active hexaaminobenzene connected by transition metal atoms. In Table 1, lattice constants of energy minimized structures along with the important bond lengths in these systems are summarized. As it is clear, the variations in the C-N and C-C bond lengths in different structures are negligible, and maximum differences are 2% and 4%, for C-N and C-C bond lengths, respectively. This way, the differences in the lattice constants of different structures originate mainly from the transition metal atoms-N bonds. In Table 1, the lattice energies of studied monolayers are also compared. For the convenience of the future studies, the unit-cells of energy minimized structures are all given in the supplementary information document.

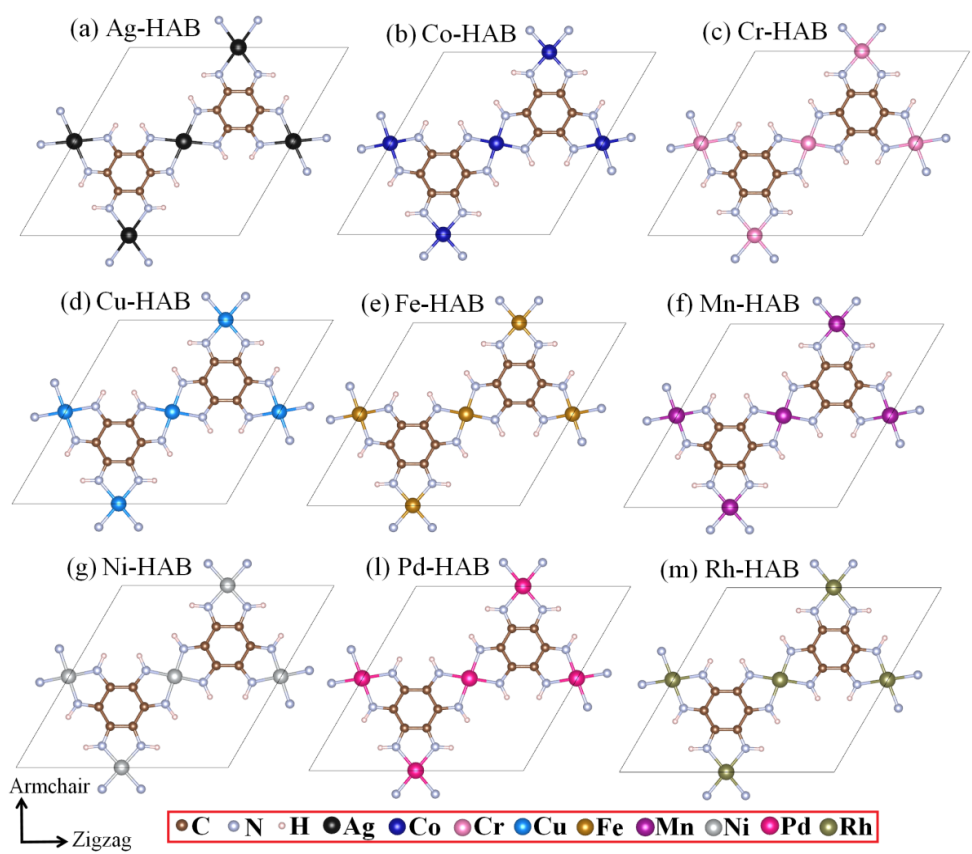

Fig. 1, Atomic structure of energy minimized single-layer Ag-, Co-, Cr-, Cu-, Fe-, Mn-, Ni-, Pd- and Rh-HAB.

Table. 1, Lattice parameters of energy minimized metal-HAB monolayers. Here, $L_{M-N}$, $L_{N-C}$ and $L_{C-C}$, and $E_{unit-cell}$ stand for metal-N, N-C and C-C bond lengths and energy per atom of a unit-cell, respectively.

| Structure | Lattice constant (Å) | $L_{M-N}$ (Å) | $L_{N-C}$ (Å) | $L_{C-C}$ (Å) | $E_{unit-cell}$ (eV/atom) |
|---|---|---|---|---|---|
| Ag-HAB | 14.87 | 2.27 | 1.31 | 1.46 | -6.518 |
| Co-HAB | 13.40 | 1.84 | 1.35 | 1.44 | -7.008 |
| Cr-HAB | 13.88 | 1.98 | 1.35 | 1.44 | -7.228 |
| Cu-HAB | 13.78 | 1.98 | 1.33 | 1.45 | -6.653 |
| Fe-HAB | 13.52 | 1.88 | 1.35 | 1.44 | -7.095 |
| Mn-HAB | 13.67 | 1.92 | 1.36 | 1.43 | -7.192 |
| Ni-HAB | 13.37 | 1.84 | 1.35 | 1.43 | -6.886 |
| Pd-HAB | 13.89 | 2.00 | 1.34 | 1.44 | -6.806 |
| Rh-HAB | 13.93 | 2.00 | 1.34 | 1.45 | -6.984 |

We next explore the mechanical responses of HAB MOFs monolayers on the basis of uniaxial tensile simulations results. In order to investigate the anisotropy in the mechanical properties, uniaxial tensile simulations were conducted along the armchair and zigzag directions, in analogy to graphene (as specified in Fig. 1). For the uniaxial tensile simulations, the periodic simulation box size along the loading direction was increased gradually with a fixed engineering strain increment. To ensure the satisfaction of uniaxial stress-conditions, the simulation box size along the sheet perpendicular direction of loading was altered to reach a negligible stress [32]. According to our results, Ag- and Cu-HAB were found to exhibit elastic instability with Poisson's ratios over 0.5, so we could not evaluate their mechanical properties. On the other side, the rest of considered nanosheets show elastic stability, with very similar trends in their uniaxial stress-strain curves. In Fig. 2, the first-principles results for the uniaxial stress-strain responses of single-layer Co-, Rh- and Pd-HAB elongated along the zigzag and armchair directions are compared. As it is clear, all predicted uniaxial stress-strain curves clearly exhibit initial linear responses, corresponding to the linear elasticity and elastic modulus. Observation of linear elasticity confirms that despite of the porous atomic lattices of transition metal-HAB monolayers, from the early stages of loading their deformation are mainly achieved by the bond-elongation rather than structural deflection. Moreover, for the uniaxial loading along the armchair and zigzag directions, the stress-strain relations match very closely initially, thus suggesting the convincingly isotropic elastic responses along these monolayers. Within the elastic range and at initial strain levels, the strain along the traverse direction of loading ($\varepsilon_t$) with respect to the loading strain ($\varepsilon_l$) is constant and can be used to evaluate the Poisson's ratio, using: $-\varepsilon_t/\varepsilon_l$ [32].

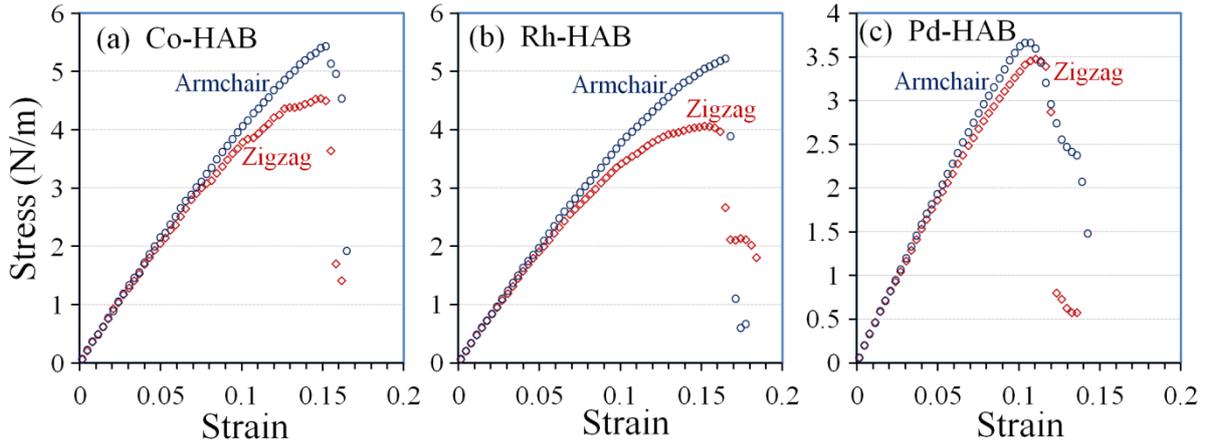

Fig. 2, Uniaxial stress-strain responses of single-layer Co-, Rh- and Pd-HAB elongated along the armchair and zigzag directions.

Another important mechanical characteristics of a material are the ultimate tensile strength and its corresponding strain, which defines the stretchability of a material. According to our results, all studied monolayers were found to yield higher tensile strengths along the armchair direction than the zigzag direction. Nevertheless, stretchability of these monolayers are considerably close for different loading directions. In Table 2, the mechanical properties of Co-, Cr-, Fe-, Mn-, Ni-, Pd- and Rh-HAB monolayers along the armchair and zigzag directions are summarized. As it is clear, among the studied nanomembranes, Co-HAB is the strongest material and Cr-HAB when stretched along the zigzag shows the minimum rigidity and strength as well. In addition, Cr-HAB and Rh-HAB exhibit the highest anisotropy ratio in tensile strengths, in contrast with Mn-HAB and Pd-HAB nanosheets in which the tensile strengths are very close along different loading directions. Interestingly, despite of the porous atomic lattices in metal-HAB nanosheets, their elastic modulus and tensile strengths are comparable or even higher than some other densely packed 2D materials, like; germanene and stanene [41].

To better understand the failure mechanism of considered metal-HAB nanosheets, we consider the structures after the tensile strength point. In Fig. 3 top views of Co-, Fe-, Mn- and Pd-HAB monolayers elongated along the armchair and zigzag directions are illustrated. We found that for the all monolayers along the both loading directions, the failure always initiates by the breakage of metal-N bonds. This observation highlights that the elongation of metal-N bonds dominates the mechanical responses of metal-HAB monolayers. During the uniaxial loading, basically the bonds oriented along the loading direction stretch, whereas as a results of the sheet shrinkage along the width, the lengths of bonds oriented along the perpendicular direction of loading may contract or only slightly change.

Table 2, Summarized mechanical properties of single-layer Co-, Cr-, Fe-, Mn-, Ni-, Pd- and Rh-HAB elongated along the armchair and zigzag directions. $E$, $P$, $UTS$ and $SUTS$ stand for elastic modulus, Poisson's ratio, ultimate tensile strength and strain at ultimate tensile strength point, respectively.

| Structure | Direction | E (N/m) | P | UTS (N/m) | SUTS |
|---|---|---|---|---|---|
| Co-HAB | Armchair | 44 | 0.40 | 5.44 | 0.15 |
| | Zigzag | 45 | 0.42 | 4.53 | 0.15 |
| Cr-HAB | Armchair | 37 | 0.38 | 4.51 | 0.18 |
| | Zigzag | 37 | 0.4 | 3.39 | 0.16 |
| Fe-HAB | Armchair | 44 | 0.40 | 5.20 | 0.15 |
| | Zigzag | 44 | 0.42 | 4.42 | 0.15 |
| Mn-HAB | Armchair | 42 | 0.37 | 4.10 | 0.11 |
| | Zigzag | 42 | 0.39 | 4.00 | 0.12 |
| Ni-HAB | Armchair | 43 | 0.42 | 3.60 | 0.09 |
| | Zigzag | 43 | 0.43 | 3.52 | 0.10 |
| Pd-HAB | Armchair | 42 | 0.38 | 3.66 | 0.11 |
| | Zigzag | 42 | 0.38 | 3.49 | 0.11 |
| Rh-HAB | Armchair | 43 | 0.38 | 5.22 | 0.16 |
| | Zigzag | 43 | 0.4 | 4.07 | 0.15 |

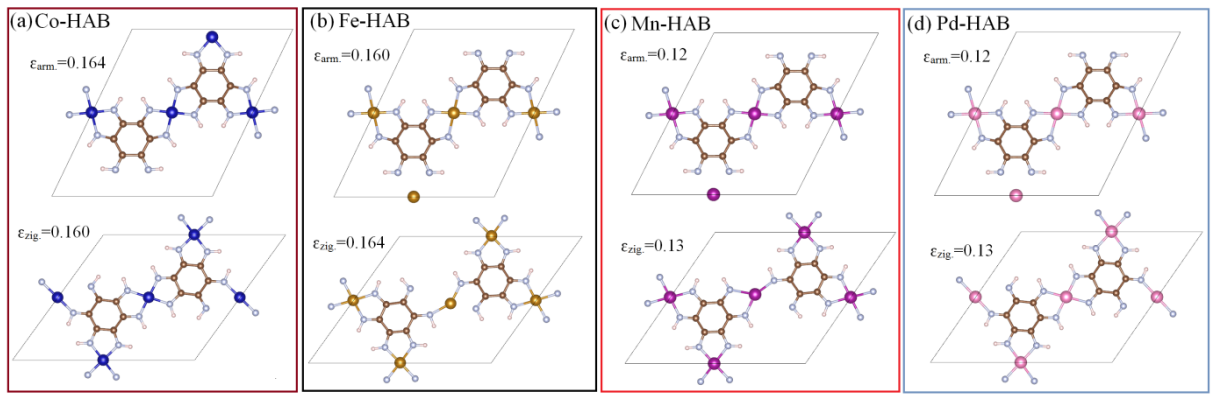

Fig. 3, Top views of the single-layer Co-, Fe-, Mn- and Pd-HAB at strain levels after the ultimate tensile strength point. $\varepsilon_{arm}$ and $\varepsilon_{zig}$ depict the engineering strain levels for the uniaxial loading along the armchair and zigzag directions, respectively.

To investigate the thermal stability of these novel 2D systems, we conducted the AIMD calculations at 1000 K. The simulations were performed for 15 ps and top and side views of the studied nanomembranes are depicted in Fig. 4. Interestingly, as it can be concluded from the results shown in Fig. 4, all considered nanosheets could stay intact at the high temperature of 1000 K, which confirm their remarkable thermal stability. Nevertheless, our simulations at the higher temperature of 2000 K, reveal the fast disintegration of these nanosheets.

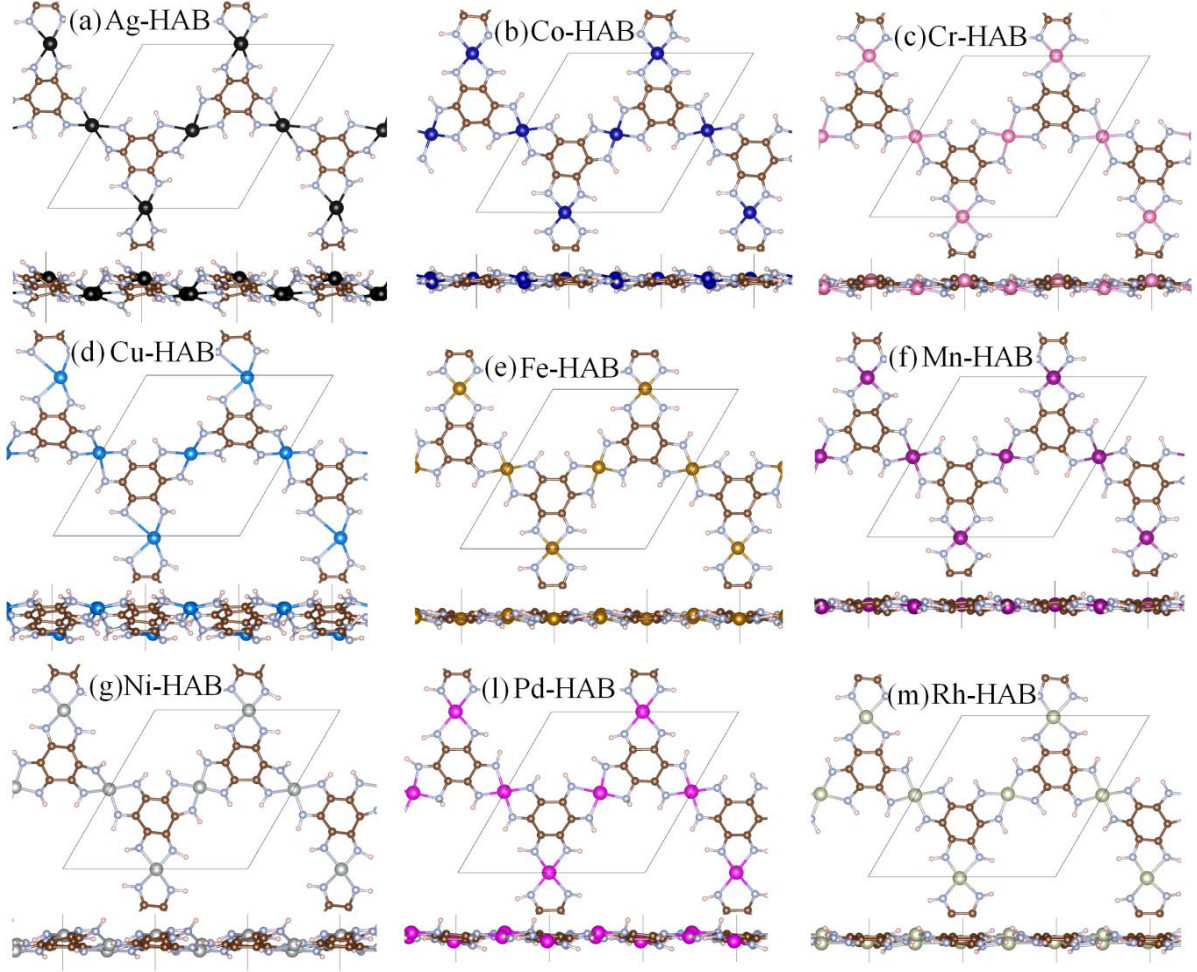

Fig. 4, Top and side views of HAB monolayers at 1000 K after the AIMD simulations for 15 ps.

To explore electrical features of transition metal-HAB nanosheets, the density of states (DOS) were calculated and shown in Fig. 5. For Ag-, Cu-, Cr-, Co-, Mn- and Fe-HAB monolayers spin splitting of DOS around Fermi energy is observable due to the strong hybridization of transition metal 3d orbital with 2s and 2p orbitals of nitrogen atoms. For these nanosheets, our calculations show the ferromagnetic (FM) phase is more stable than nonmagnetic (NM) phase. The spin splitting in these systems leads to spin polarization and magnetization which are given in Table 3. The magnetization, $M$, and the spin polarization, $P$, are defined as follows:

$$M = (N^{\uparrow} - N^{\downarrow}) \times \mu_B \quad (1)$$

$$P = \frac{N_{Ef}^{\uparrow} - N_{Ef}^{\downarrow}}{N_{Ef}^{\uparrow} + N_{Ef}^{\downarrow}} \quad (2)$$

where $N^{\uparrow}$, $N^{\downarrow}$, $N_{Ef}^{\uparrow}$ and $N_{Ef}^{\downarrow}$ represent the total number of electrons of majority spin, total number of electrons of minority spin, the density of states of majority spin (spin up) and minority spin (spin down) at the Fermi level, respectively. Our results show that the Ag-, Cu-, Cr and Mn-HAB monolayers are half-metallic ferromagnetic systems with 100% spin polarization, because they act as a conductor to electrons of one spin orientation but as a semiconductor to those of the opposite orientation. In these half-metal systems, the valence band for one spin orientation is partially filled, while there is a gap in the density of states for the other spin orientation. We have also calculated entire system magnetic moment and partial magnetic moments of each magnetic atoms. The total magnetic moment per magnetic atoms is 0.32, 0.11, 2.69 and 2.86 $\mu_B$ for the Ag-, Cu-, Cr and Mn-HAB monolayers, respectively. The spin-flip gap which is defined as the separation of the Fermi level from the minimum of the conduction band in minority spin was estimated to be 0.57, 0.00, 0.00 and 0.17 for Ag-, Cu-, Cr- and Mn-HAB nanosheets. This parameter is important for the electron injection in half metals. The Co- and Fe-HAB monolayers show metallic features for the both spin channels with 36% and 12.5% spin polarization. The total magnetic moment for these systems is 0.34 and 1.88 $\mu_B$. The high magnetic moment and spin polarization of these systems imply that they can be used in electronic devices, magnetic recording and spintronics. The Ni-, Pd- and Rh-HAB systems are nonmagnetic metals since there is no spin splitting between spin up and spin down of electronic states.

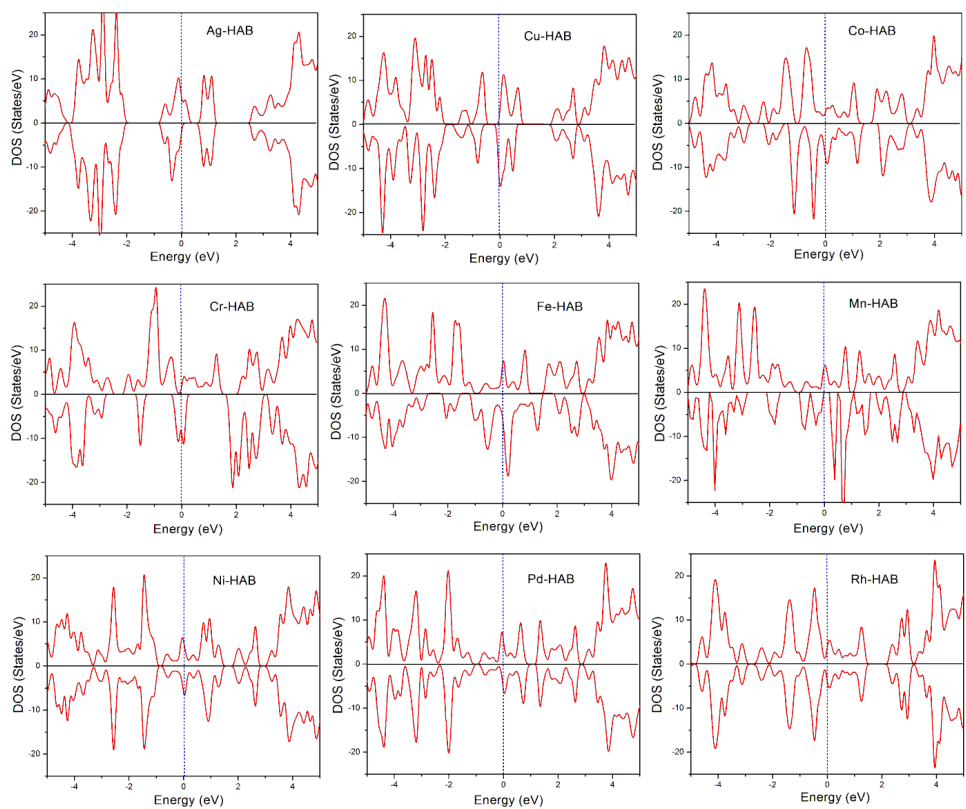

Fig. 5, Total electronic density of states.

Table 3, Summarized electronic properties of metal-HAB monolayers. Here, TMM and PMM depict the total magnetic moments per magnetic atom and partial magnetic moments of the magnetic atom, respectively.

| Structure | Polarization (%) | TMM | PMM |
|---|---|---|---|
| Cr-HAB | 100 | 2.69 | 3.12 |
| Mn-HAB | 100 | 2.86 | 3.17 |
| Fe-HAB | 12.50 | 1.88 | 1.99 |
| Co-HAB | 36 | 0.34 | 0.42 |
| Ni-HAB | 0.00 | 0.00 | 0.00 |
| Pd-HAB | 0.00 | 0.00 | 0.00 |
| Rh-HAB | 0.00 | 0.00 | 0.00 |
| Cu-HAB | 100 | 0.11 | 0.44 |
| Ag-HAB | 100 | 0.32 | 0.17 |

## 4. Concluding remarks

In recent experimental advances, hexaaminobenzene (HAB)-derived two-dimensional metal–organic frameworks were synthesized, with excellent application prospects for advanced energy systems. Motivated by successful synthesis of Ni-, Co- and Cu-HAB nanosheets, in this short communication we also predicted Ag-, Cr-, Fe-, Mn-, Pd- and Rh-HAB nanosheets, with similar atomic lattices as those of experimentally realized nanomembranes. We then employed first-principles density functional theory calculations to obtain the energy minimized atomic lattices and investigate mechanical properties, thermal stability, and electronic responses of free-standing and single-layer Ag-, Co-, Cr-, Cu-, Fe-, Mn-, Ni-, Pd- and Rh-HAB. Mechanical/failure responses of aforementioned monolayers were analyzed by performing uniaxial tensile simulations. It was confirmed that despite of porous atomic lattices, Co-, Cr-, Fe-, Mn-, Ni-, Pd- and Rh-HAB nanosheets exhibit linear elasticity with considerable tensile strengths. Notably, despite of porous atomic lattices, elastic modulus and tensile strengths of aforementioned were found to be comparable or even higher than some other densely packed 2D materials. AIMD simulations highlight remarkable thermal stability of all studied monolayers, with the intact atomic lattices at the high temperature of 1000 K. Half-metallic properties for the Ag-, Cu-, Cr-, and Mn-HAB monolayers were predicted within the PBE approach. The corresponding total magnetic moment per magnetic atoms for aforementioned monolayers was predicted to be 0.32, 0.11, 2.69 and 2.86 $\mu_B$, respectively. The electronic and magnetic calculations show that Co- and Fe-HAB monolayers are metals with a very low spin polarization at the Fermi level and magnetic moments of 0.34 and 1.88 $\mu_B$, respectively. The high magnetic moment and spin polarization of studied nanosheets imply that they can be used in electronic devices and spintronics. It was moreover observed that Ni-, Pd- and Rh- HAB monolayers exhibit nonmagnetic metallic behavior.

**Acknowledgment**

B. M. and T. R. greatly acknowledge the financial support by European Research Council for COMBAT project (Grant number 615132).

**Data availability**

The energy minimized atomic lattices are available to download from:

# First-principles investigation of Ag-, Co-, Cr-, Cu-, Fe-, Mn-, Ni-, Pd- and Rh-hexaaminobenzene 2D metal-organic frameworks

Bohayra Mortazavi[*,1], Masoud Shahrokhi[1], Meysam Makaremi[2], Gianaurelio Cuniberti[3,4,5] and Timon Rabczuk[6]

[1]*Institute of Structural Mechanics, Bauhaus-Universität Weimar, Marienstr. 15, D-99423 Weimar, Germany.*

[2]*Department of Materials Science and Engineering, University of Toronto, 184 College Street, Suite 140, Toronto, ON M5S 3E4, Canada.*

[3]*Institute for Materials Science and Max Bergman Center of Biomaterials, TU Dresden, 01062 Dresden, Germany*

[4]*Dresden Center for Computational Materials Science, TU Dresden, D-01062 Dresden, Germany*

[5]*Center for Advancing Electronics Dresden, TU Dresden, 01062 Dresden, Germany*

[6]*Department of Computer Engineering, College of Computer and Information Sciences, King Saud University, Riyadh, Saudi Arabia.*

*E-mail: bohayra.mortazavi@gmail.com

1. Atomic structures of constructed monolayers unit-cells in VASP POSCAR

# 1.1 Ag-HAB

```
C4N4H4Ag
   1.00000000000000
     14.8688789099722705    0.0000000000000000    0.0000000000000000
      7.4344394550419066   12.8768268618653998    0.0000000000000000
      0.0000000000000000    0.0000000000000000   20.0000000000000000
   C    N    H    Ag
   12   12   12    3
Direct
  0.3302373961154288  0.2361700066310775  0.5000000000000000
  0.4289383331841705  0.2361396541869425  0.5000000000000000
  0.4289733006555778  0.3345156863868581  0.5000000000000000
  0.3303490811216108  0.4332679670836512  0.5000000000000000
  0.2320103932788129  0.4332463772745783  0.5000000000000000
  0.2319674833795844  0.3346059009826661  0.5000000000000000
  0.7628695505248118  0.5690136981675948  0.5000000000000000
  0.7628042536647186  0.6677237308675572  0.5000000000000000
  0.6644399457529246  0.7660754815166158  0.5000000000000000
  0.5658064537489338  0.7659604426262518  0.5000000000000000
  0.5658926832268207  0.6675609872717843  0.5000000000000000
  0.6645806892155548  0.5688999791709719  0.5000000000000000
  0.3361237308534442  0.1450856435785184  0.5000000000000000
  0.5140472336992019  0.1450003352594180  0.5000000000000000
  0.5141505332248286  0.3404358627567063  0.5000000000000000
  0.3363579509487948  0.5183937915868384  0.5000000000000000
  0.1408698211003028  0.5184129286274484  0.5000000000000000
  0.1408038417643098  0.3406355806920445  0.5000000000000000
  0.8540211054267743  0.4838712205133291  0.5000000000000000
  0.8539208885494887  0.6617818239427322  0.5000000000000000
  0.6583932554677006  0.8572248306232240  0.5000000000000000
  0.4806063766252393  0.8570338796513681  0.5000000000000000
  0.4807842709906609  0.6615057022494426  0.5000000000000000
  0.6586285414434698  0.4837302523983240  0.5000000000000000
  0.2655295226334928  0.1488019376185505  0.5000000000000000
  0.5809831647117036  0.1485862824486262  0.5000000000000000
  0.5810519302366828  0.2698792730330908  0.5000000000000000
  0.2657913775327998  0.5853105085363026  0.5000000000000000
  0.1443818840080411  0.5853835088433428  0.5000000000000000
  0.1442629594864329  0.2701755536681318  0.5000000000000000
  0.8505141131472911  0.4168932738679985  0.5000000000000000
  0.8503448031085057  0.7322941383165826  0.5000000000000000
  0.7289306306394607  0.8536255113996418  0.5000000000000000
  0.4137464400708026  0.8533000388118310  0.5000000000000000
  0.4138178046118242  0.7320061099719410  0.5000000000000000
  0.7291739368193362  0.4167948532387555  0.5000000000000000
  0.4971996727459924  0.0010651594151554  0.5000000000000000
  0.4975551264196127  0.5009319063901216  0.5000000000000000
  0.9974114538649488  0.5011963423939392  0.5000000000000000
```

## 1.2 Co-HAB

```
C4N4H4Co
1.00000000000000
    13.4077726866508495     0.0000000000000000     0.0000000000000000
     6.7038863433757188    11.6114717548381297     0.0000000000000000
     0.0000000000000000     0.0000000000000000    20.0000000000000000
   C    N    H    Co
   12   12   12    3
Direct
  0.3301954114370318  0.2277090292840285  0.5000000000000000
  0.4378573131897454  0.2277011352738967  0.5000000000000000
  0.4379052590040686  0.3339355080617281  0.5000000000000000
  0.3302782656136571  0.4416436342414087  0.5000000000000000
  0.2239602308109880  0.4416917740502342  0.5000000000000000
  0.2239188254201423  0.3340126266060679  0.5000000000000000
  0.7708539325946262  0.5605592692117440  0.5000000000000000
  0.7708224567207225  0.6682469517809793  0.5000000000000000
  0.6645267341570715  0.7745027359875820  0.5000000000000000
  0.5568586201100558  0.7744771856684025  0.5000000000000000
  0.5569165430877661  0.6681447319493188  0.5000000000000000
  0.6646058951491369  0.5604868151538156  0.5000000000000000
  0.3468000315624948  0.1200550248375905  0.5000000000000000
  0.5288796742291950  0.1200381149796286  0.5000000000000000
  0.5289977592500179  0.3504812746029273  0.5000000000000000
  0.3469736029426400  0.5326370146464754  0.5000000000000000
  0.1163121654787105  0.5327377934246087  0.5000000000000000
  0.1162591904423778  0.3506702577786953  0.5000000000000000
  0.8785047553784437  0.4695469134549626  0.5000000000000000
  0.8784564835367874  0.6516615042698035  0.5000000000000000
  0.6478377445867380  0.8821923373829037  0.5000000000000000
  0.4658009657593780  0.8821198485741633  0.5000000000000000
  0.4659048648094171  0.6514801639640950  0.5000000000000000
  0.6480267282253180  0.4694193115114160  0.5000000000000000
  0.2742098404232181  0.1133519379888739  0.5000000000000000
  0.6081712204370575  0.1133304307942922  0.5000000000000000
  0.6082534861878557  0.2778516511833331  0.5000000000000000
  0.2744453696720939  0.6119790112795584  0.5000000000000000
  0.1095608901114176  0.6120534542133882  0.5000000000000000
  0.1094861270677754  0.2781188861868316  0.5000000000000000
  0.8852322128467023  0.3902419150278149  0.5000000000000000
  0.8851511145036426  0.7242606051765961  0.5000000000000000
  0.7203833488083404  0.8889702836698916  0.5000000000000000
  0.3865064594098229  0.8888278541839227  0.5000000000000000
  0.3865749392447952  0.7240267675808951  0.5000000000000000
  0.7206261007003150  0.3901330987961060  0.5000000000000000
  0.4973216533974139  0.0011008163231736  0.5000000000000000
  0.4975065039348650  0.5009744758829910  0.5000000000000000
  0.9973892137582183  0.5011640170158458  0.5000000000000000
```

## 1.3 Cr-HAB

```
C4N4H4Cr
1.00000000000000
    13.8789753111386904     0.0000000000000000     0.0000000000000000
     6.9394876556213880    12.0195451979755603     0.0000000000000000
     0.0000000000000000     0.0000000000000000    20.0000000000000000
     C    N    H   Cr
    12   12   12    3
Direct
  0.3302481736132327  0.2311935990593188  0.5000000000000000
  0.4343791567533280  0.2311423249730496  0.5000000000000000
  0.4344196497587646  0.3339755595824876  0.5000000000000000
  0.3303193470822345  0.4381144158334073  0.5000000000000000
  0.2274642500289843  0.4381445816058059  0.5000000000000000
  0.2274260029145349  0.3340492179714903  0.5000000000000000
  0.7673952702910660  0.5640448923335407  0.5000000000000000
  0.7673488266222077  0.6681995507871084  0.5000000000000000
  0.6644846242432446  0.7710254044419633  0.5000000000000000
  0.5603993531656043  0.7709746607728657  0.5000000000000000
  0.5604328494753090  0.6681157039571676  0.5000000000000000
  0.6645659884813743  0.5640096135368747  0.5000000000000000
  0.3429279736674498  0.1280568444646321  0.5000000000000000
  0.5247639886988151  0.1279693307429889  0.5000000000000000
  0.5248505117351473  0.3466824197076284  0.5000000000000000
  0.3430880013384945  0.5285055299898802  0.5000000000000000
  0.1243303288266646  0.5285809739264726  0.5000000000000000
  0.1242653887548977  0.3468216906875816  0.5000000000000000
  0.8705552629741002  0.4736496502527826  0.5000000000000000
  0.8704812757472474  0.6555244699636579  0.5000000000000000
  0.6516686967237865  0.8742191862795031  0.5000000000000000
  0.4699287516069614  0.8741158872720299  0.5000000000000000
  0.4700082649208508  0.6553759763321046  0.5000000000000000
  0.6518272814525070  0.4735862604457846  0.5000000000000000
  0.2703116513410149  0.1262452332361057  0.5000000000000000
  0.5992376975500372  0.1260602354945445  0.5000000000000000
  0.5992917520120924  0.2740731902635209  0.5000000000000000
  0.2705102652190198  0.6029715026704614  0.5000000000000000
  0.1224969276948707  0.6030209105145597  0.5000000000000000
  0.1223611134850741  0.2742525877544526  0.5000000000000000
  0.8724191167902048  0.3992016652997291  0.5000000000000000
  0.8722684667319539  0.7281485115006208  0.5000000000000000
  0.7242208878523387  0.8761441699077708  0.5000000000000000
  0.3955083838167845  0.8759182876603738  0.5000000000000000
  0.3955596217697019  0.7279508150731794  0.5000000000000000
  0.7244159520067835  0.3991500564405828  0.5000000000000000
  0.4972602109046235  0.0010925058188187  0.5000000000000000
  0.4974344770219047  0.5010516568015750  0.5000000000000000
  0.9973961909268587  0.5011770886436394  0.5000000000000000
```

## 1.4 Cu-HAB

```
C4N4H4Cu
1.00000000000000
     13.7826922150136202    0.0000000000000000    0.0000000000000000
      6.8913461075585252   11.9361615907760292    0.0000000000000000
      0.0000000000000000    0.0000000000000000   20.0000000000000000
    C    N    H    Cu
    12   12   12    3
Direct
  0.3304575208233604  0.2294403345809980  0.5000000000000000
  0.4359920122288362  0.2293970677208477  0.5000000000000000
  0.4360420801616719  0.3341479974115302  0.5000000000000000
  0.3305379752982769  0.4396444820311991  0.5000000000000000
  0.2257340437378365  0.4397052436525989  0.5000000000000000
  0.2256864098804357  0.3342139518390925  0.5000000000000000
  0.7690889605304825  0.5625447544180976  0.5000000000000000
  0.7690660335020607  0.6680489542512345  0.5000000000000000
  0.6642433363993181  0.7727916502227608  0.5000000000000000
  0.5587390416608685  0.7727765355792366  0.5000000000000000
  0.5587808341755096  0.6679519417232377  0.5000000000000000
  0.6643118494139725  0.5624817907801765  0.5000000000000000
  0.3411370309954260  0.1281821456496071  0.5000000000000000
  0.5265436607969203  0.1281288678497364  0.5000000000000000
  0.5266158630120348  0.3448202976694223  0.5000000000000000
  0.3412435496596157  0.5302025614163908  0.5000000000000000
  0.1244880903186111  0.5302692898580734  0.5000000000000000
  0.1244338013413867  0.3449098337575717  0.5000000000000000
  0.8703359030100463  0.4719846684511868  0.5000000000000000
  0.8703120762939921  0.6573759370083412  0.5000000000000000
  0.6535468809280819  0.8740497289744866  0.5000000000000000
  0.4681745074369863  0.8740192555506473  0.5000000000000000
  0.4682330988157726  0.6572394607066786  0.5000000000000000
  0.6536548436831551  0.4718999339912702  0.5000000000000000
  0.2677840564994241  0.1267447822766735  0.5000000000000000
  0.6013448938959238  0.1266775708831324  0.5000000000000000
  0.6013954340567004  0.2714489892871896  0.5000000000000000
  0.2678926951801568  0.6049912433896280  0.5000000000000000
  0.1230702371442194  0.6050501298925079  0.5000000000000000
  0.1229893470997410  0.2715547227917848  0.5000000000000000
  0.8717475264172836  0.3971966606615412  0.5000000000000000
  0.8716899955944336  0.7307648201930850  0.5000000000000000
  0.7269029299491426  0.8754939209914987  0.5000000000000000
  0.3933987722236747  0.8754194314787561  0.5000000000000000
  0.3934365833047792  0.7305964768076515  0.5000000000000000
  0.7270228101189485  0.3971202674892357  0.5000000000000000
  0.4973434878559825  0.0010880835720795  0.5000000000000000
  0.4974525353309966  0.5010242928479192  0.5000000000000000
  0.9974012252239990  0.5011380843428852  0.5000000000000000
```

## 1.5 Fe-HAB

```
C4N4H4Fe
1.00000000000000
    13.5242559248653400     0.0000000000000000     0.0000000000000000
     6.7621279624834019    11.7123491982472192     0.0000000000000000
     0.0000000000000000     0.0000000000000000    20.0000000000000000
   C    N    H    Fe
   12   12   12    3
Direct
  0.3302658576262587  0.2288360212967504  0.5000000000000000
  0.4366533968850348  0.2288130976123099  0.5000000000000000
  0.4366816942339398  0.3340314367146746  0.5000000000000000
  0.3303377319455123  0.4404353730698531  0.5000000000000000
  0.2250927457645489  0.4404740950974286  0.5000000000000000
  0.2250597338066456  0.3340940069931335  0.5000000000000000
  0.7696866000410125  0.5617847414077417  0.5000000000000000
  0.7696359371338062  0.6681743960066271  0.5000000000000000
  0.6643837440658018  0.7734057408184754  0.5000000000000000
  0.5579909860029204  0.7734099810776911  0.5000000000000000
  0.5580571080731929  0.6681085417889676  0.5000000000000000
  0.6644804197165897  0.5617345446631745  0.5000000000000000
  0.3457996706138644  0.1218464330687752  0.5000000000000000
  0.5281269960998856  0.1218394942224705  0.5000000000000000
  0.5282132741555188  0.3494494808713142  0.5000000000000000
  0.3459840546604553  0.5318272826104788  0.5000000000000000
  0.1181233747389498  0.5319435990181987  0.5000000000000000
  0.1180788652294069  0.3496444061048365  0.5000000000000000
  0.8766851688081090  0.4703480904905035  0.5000000000000000
  0.8766174657424131  0.6526773286988359  0.5000000000000000
  0.6488755783683260  0.8803795405322745  0.5000000000000000
  0.4665364080736580  0.8803784835699631  0.5000000000000000
  0.4666646391623814  0.6524884189779456  0.5000000000000000
  0.6490471309227388  0.4702196288204021  0.5000000000000000
  0.2729237694243380  0.1167787657610120  0.5000000000000000
  0.6060589260969707  0.1168104528979299  0.5000000000000000
  0.6061010862063156  0.2764872307080637  0.5000000000000000
  0.2731619895826825  0.6098181522960786  0.5000000000000000
  0.1131115763950419  0.6098608568718760  0.5000000000000000
  0.1130449367512085  0.2767513182855339  0.5000000000000000
  0.8817704022752366  0.3923986673933655  0.5000000000000000
  0.8816735087881042  0.7255581478977220  0.5000000000000000
  0.7217750498824742  0.8854131850174056  0.5000000000000000
  0.3886107645115544  0.8854065170128820  0.5000000000000000
  0.3886981929422859  0.7253453751819450  0.5000000000000000
  0.7220156889439977  0.3923345978210975  0.5000000000000000
  0.4973437738295630  0.0011134770450525  0.5000000000000000
  0.4975269297101761  0.5009463911431453  0.5000000000000000
  0.9973767567891088  0.5011688631341045  0.5000000000000000
```

## 1.6 Mn-HAB

```
C4N4H4Mn
1.00000000000000
     13.6690926074036696     0.0000000000000000     0.0000000000000000
      6.8345463037530934    11.8377814447255805     0.0000000000000000
      0.0000000000000000     0.0000000000000000    20.0000000000000000
   C    N    H    Mn
   12   12   12    3
Direct
  0.3303915380325435  0.2302661622684994  0.5000000000000000
  0.4352165897603015  0.2302017348066840  0.5000000000000000
  0.4352765915837296  0.3340614096133550  0.5000000000000000
  0.3305024628139464  0.4388625362237732  0.5000000000000000
  0.2265534420282833  0.4389448769252411  0.5000000000000000
  0.2264980365937888  0.3341903652423213  0.5000000000000000
  0.7682306642038341  0.5633177257816016  0.5000000000000000
  0.7681866906684931  0.6681115567686717  0.5000000000000000
  0.6642755856426561  0.7719826156217309  0.5000000000000000
  0.5595055864806184  0.7719388487310326  0.5000000000000000
  0.5595562246207280  0.6679807005297178  0.5000000000000000
  0.6643768971105430  0.5632277305620619  0.5000000000000000
  0.3444322574199674  0.1242007935737135  0.5000000000000000
  0.5271865911328462  0.1241099783223234  0.5000000000000000
  0.5272982800221016  0.3480934601828594  0.5000000000000000
  0.3446357234873787  0.5308255258992247  0.5000000000000000
  0.1204939078934473  0.5309825414203644  0.5000000000000000
  0.1204043593084450  0.3483107985513172  0.5000000000000000
  0.8743406726767944  0.4713560107443087  0.5000000000000000
  0.8742531966900842  0.6540792785949634  0.5000000000000000
  0.6501875639355035  0.8780635562507868  0.5000000000000000
  0.4674957615015742  0.8780105720178000  0.5000000000000000
  0.4675974962131804  0.6538597278461751  0.5000000000000000
  0.6503872831236706  0.4711844447381210  0.5000000000000000
  0.2722812719235463  0.1196395519916180  0.5000000000000000
  0.6039224651632651  0.1194941980369890  0.5000000000000000
  0.6040044859944587  0.2759327928520534  0.5000000000000000
  0.2725319142863327  0.6075744409959185  0.5000000000000000
  0.1159470268635019  0.6076889167013348  0.5000000000000000
  0.1157899521392380  0.2761895020847973  0.5000000000000000
  0.8789796425067280  0.3946089713277345  0.5000000000000000
  0.8788021804819124  0.7262397142029471  0.5000000000000000
  0.7223190086066182  0.8826590780438082  0.5000000000000000
  0.3907729947139984  0.8825969010574752  0.5000000000000000
  0.3908644873707985  0.7259856864934804  0.5000000000000000
  0.7225699911484611  0.3944978712359344  0.5000000000000000
  0.4973045942542242  0.0010978008687772  0.5000000000000000
  0.4975223051329181  0.5009522514109150  0.5000000000000000
  0.9973762104696036  0.5012155334796660  0.5000000000000000
```

## 1.7 Ni-HAB

```
C4N4H4Ni
1.00000000000000
     13.3727099594965306     0.0000000000000000     0.0000000000000000
      6.6863549797984287    11.5811065423964603     0.0000000000000000
      0.0000000000000000     0.0000000000000000    20.0000000000000000
   C    N    H    Ni
    12   12   12    3
Direct
  0.3305372707180277  0.2273786450692038  0.5000000000000000
  0.4378109342140571  0.2273897745806910  0.5000000000000000
  0.4378673132452318  0.3342587591939590  0.5000000000000000
  0.3306244649831911  0.4416015359103145  0.5000000000000000
  0.2236655415181646  0.4416573641911565  0.5000000000000000
  0.2236346972162766  0.3343281377587330  0.5000000000000000
  0.7711426139649079  0.5606154449369738  0.5000000000000000
  0.7711132808231463  0.6679197209873974  0.5000000000000000
  0.6641716744546424  0.7748347426840805  0.5000000000000000
  0.5568963972705704  0.7747958116024194  0.5000000000000000
  0.5569330593554823  0.6678242535547142  0.5000000000000000
  0.6642545608122110  0.5605326114805781  0.5000000000000000
  0.3463551216883751  0.1197338149524208  0.5000000000000000
  0.5296273822664008  0.1197443374890485  0.5000000000000000
  0.5297340917708147  0.3500105635535271  0.5000000000000000
  0.3465406764603620  0.5333661056490863  0.5000000000000000
  0.1160190454463569  0.5334745524358531  0.5000000000000000
  0.1159714494711147  0.3502139832581228  0.5000000000000000
  0.8787887951092139  0.4688135261155560  0.5000000000000000
  0.8787507980281594  0.6521175942859014  0.5000000000000000
  0.6482888053947587  0.8825057047446163  0.5000000000000000
  0.4650540528602720  0.8824310594571394  0.5000000000000000
  0.4651548614908053  0.6519362019055532  0.5000000000000000
  0.6484820167340833  0.4686829093951268  0.5000000000000000
  0.2747694826026040  0.1113279600168643  0.5000000000000000
  0.6096105860555724  0.1113575003941682  0.5000000000000000
  0.6096958648958446  0.2783710741345834  0.5000000000000000
  0.2750258802362859  0.6134180589931204  0.5000000000000000
  0.1075660181794056  0.6134881180569370  0.5000000000000000
  0.1075054818372088  0.2786661926058015  0.5000000000000000
  0.8871900624955875  0.3888236213328073  0.5000000000000000
  0.8871348966305891  0.7237144339563771  0.5000000000000000
  0.7198400619510584  0.8909674159515599  0.5000000000000000
  0.3850606364850908  0.8908351561934956  0.5000000000000000
  0.3851176635525700  0.7234719640704141  0.5000000000000000
  0.7201022878124178  0.3887043065586739  0.5000000000000000
  0.4973254241534010  0.0011042950045513  0.5000000000000000
  0.4975214243272986  0.5009544076665662  0.5000000000000000
  0.9973872574883897  0.5011645018719256  0.5000000000000000
```

## 1.8 Pd-HAB

```
C4N4H4Pd
1.00000000000000
     13.8907382216976796    0.0000000000000000    0.0000000000000000
      6.9453691109009412   12.0297321773421597    0.0000000000000000
      0.0000000000000000    0.0000000000000000   20.0000000000000000
    C    N    H   Pd
    12   12   12    3
Direct
  0.3306651175810347  0.2310710027763130  0.5000000000000000
  0.4340581266740755  0.2310370179136988  0.5000000000000000
  0.4341083235271380  0.3343789476741392  0.5000000000000000
  0.3307612510914177  0.4377745764483066  0.5000000000000000
  0.2273875776211466  0.4378178753030921  0.5000000000000000
  0.2273382383343874  0.3344660362361666  0.5000000000000000
  0.7674697924951389  0.5644072203606645  0.5000000000000000
  0.7674261822368157  0.6678184208440016  0.5000000000000000
  0.6640421708168481  0.7711488496369449  0.5000000000000000
  0.5606858599562782  0.7710865608709651  0.5000000000000000
  0.5607322730059244  0.6677028284100802  0.5000000000000000
  0.6641324165464852  0.5643503737568182  0.5000000000000000
  0.3419162029399017  0.1291799117623302  0.5000000000000000
  0.5246298204490683  0.1291147487501578  0.5000000000000000
  0.5247426214112000  0.3456314485143790  0.5000000000000000
  0.3421184854294765  0.5283520085684369  0.5000000000000000
  0.1254825712675078  0.5284490843598458  0.5000000000000000
  0.1254070874463147  0.3458206754618667  0.5000000000000000
  0.8693703792992293  0.4738416069625125  0.5000000000000000
  0.8693091940079420  0.6565490597659460  0.5000000000000000
  0.6526603770270631  0.8730749885652419  0.5000000000000000
  0.4700591417180204  0.8729683375179365  0.5000000000000000
  0.4701656145611928  0.6563528491900661  0.5000000000000000
  0.6528581797373979  0.4737376062924534  0.5000000000000000
  0.2710952755400868  0.1245205183710675  0.5000000000000000
  0.6001424352855360  0.1243774608205406  0.5000000000000000
  0.6002104593474588  0.2748017050307420  0.5000000000000000
  0.2713531356888607  0.6038773195479129  0.5000000000000000
  0.1207741117522569  0.6039492880199082  0.5000000000000000
  0.1206428123594843  0.2750600448350083  0.5000000000000000
  0.8740912916963942  0.3983333570611303  0.5000000000000000
  0.8739753744936828  0.7273665749132799  0.5000000000000000
  0.7234107378653937  0.8778599738942969  0.5000000000000000
  0.3945713287532290  0.8776420090991111  0.5000000000000000
  0.3946340573601810  0.7271124920524596  0.5000000000000000
  0.7236619583240276  0.3982447460396156  0.5000000000000000
  0.4972885223536925  0.0010803231367075  0.5000000000000000
  0.4975019864824887  0.5009910709887277  0.5000000000000000
  0.9973914415162716  0.5011872422470828  0.5000000000000000
```

## 1.9 Rh-HAB

```
C4N4H4Rh
1.00000000000000
    13.9338638859786794    0.0000000000000000    0.0000000000000000
     6.9669319430416250   12.0670800981647304    0.0000000000000000
     0.0000000000000000    0.0000000000000000   20.0000000000000000
   C    N    H    Rh
   12   12   12    3
Direct
  0.3303226967565478  0.2311378821547330  0.5000000000000000
  0.4343743134353631  0.2311056037510430  0.5000000000000000
  0.4343925569318611  0.3340098009047867  0.5000000000000000
  0.3303624126828449  0.4380908526875590  0.5000000000000000
  0.2274660709731326  0.4380878177616974  0.5000000000000000
  0.2274420065935817  0.3340620062005039  0.5000000000000000
  0.7673752764365903  0.5641334320596844  0.5000000000000000
  0.7673516934226031  0.6681858491983861  0.5000000000000000
  0.6644251440550031  0.7710763372583284  0.5000000000000000
  0.5604316075625614  0.7709991461180010  0.5000000000000000
  0.5604554417589335  0.6680865625189490  0.5000000000000000
  0.6644968971884441  0.5640563428109274  0.5000000000000000
  0.3431753720692988  0.1293373879864454  0.5000000000000000
  0.5232437721619665  0.1292690350212096  0.5000000000000000
  0.5233062431873811  0.3469059144544549  0.5000000000000000
  0.3433067941777588  0.5269744294575608  0.5000000000000000
  0.1256397527556737  0.5270187183605792  0.5000000000000000
  0.1255895319731246  0.3470249970309054  0.5000000000000000
  0.8692003916875920  0.4752328983219239  0.5000000000000000
  0.8691688998735208  0.6552945902540206  0.5000000000000000
  0.6514396092366511  0.8729226325372783  0.5000000000000000
  0.4714818145160653  0.8727980929534055  0.5000000000000000
  0.4715570562395831  0.6551518145796011  0.5000000000000000
  0.6515917541237561  0.4751392269308783  0.5000000000000000
  0.2714594389683525  0.1262249299456929  0.5000000000000000
  0.5981082179895812  0.1260707442541616  0.5000000000000000
  0.5981416981076118  0.2752046245882699  0.5000000000000000
  0.2716299842666945  0.6018399389723866  0.5000000000000000
  0.1224596574993313  0.6018830658261631  0.5000000000000000
  0.1223519618342363  0.2753748151788608  0.5000000000000000
  0.8723688339601042  0.4003737568600512  0.5000000000000000
  0.8722995864158234  0.7270030994265074  0.5000000000000000
  0.7230797167060948  0.8761710622896853  0.5000000000000000
  0.3966521407696660  0.8759176057376996  0.5000000000000000
  0.3966933977948623  0.7268222338271838  0.5000000000000000
  0.7232685301564559  0.4002793611769897  0.5000000000000000
  0.4973027254465139  0.0010848351579703  0.5000000000000000
  0.4974560108726223  0.5010316954214593  0.5000000000000000
  0.9974029234121827  0.5011530220239752  0.5000000000000000
```